\documentclass[iop,apj,tighten]{emulateapj}
\usepackage{apjfonts}

\usepackage{bm}
\usepackage{amsmath,amstext}
\usepackage[breaklinks,colorlinks,citecolor=blue,linkcolor=magenta]{hyperref}

\usepackage[noend]{algpseudocode}
\usepackage{algorithm}
\usepackage[all]{hypcap} 

\shortauthors{DMS}
\shorttitle{Source Reconstruction with RIM}

\usepackage{color}

\begin{document}
\title{Data-driven reconstruction of  Gravitationally Lensed Galaxies using\\ Recurrent Inference Machines}
\author{Warren R. Morningstar\altaffilmark{1}},
\author{Laurence Perreault Levasseur\altaffilmark{3}}
\author{Yashar D. Hezaveh\altaffilmark{3}}
\author{Roger Blandford\altaffilmark{1}}
\author{Phil Marshall\altaffilmark{1}}
\author{Patrick Putzky\altaffilmark{4}}
\author{Thomas D. Rueter\altaffilmark{2}}
\author{Risa Wechsler\altaffilmark{1}}
\author{Max Welling\altaffilmark{4}}

\altaffiltext{1}{Kavli Institute for Particle Astrophysics and Cosmology and Department of Physics, Stanford University, 452 Lomita Mall, Stanford, CA 94305-4085, USA}
\altaffiltext{2}{SLAC National Accelerator Laboratory and Department of Physics, Stanford University, 452 Lomita Mall, Stanford, CA 94305-4085, USA}
\altaffiltext{3}{Center for Computational Astrophysics, Flatiron Institute, 162 Fifth Avenue, New York, NY 10010, USA}
\altaffiltext{4}{Informatics Institute, University of Amsterdam, 1090 GH Amsterdam, Netherlands}

\begin{abstract}
We present a machine learning method for the reconstruction of the undistorted images of background sources in strongly lensed systems. This method treats the source as a pixelated image and utilizes the Recurrent Inference Machine (RIM) to iteratively reconstruct the background source given a lens model. Our architecture learns to minimize the likelihood of the model parameters (source pixels) given the data using the physical forward model (ray tracing simulations) while implicitly learning the prior of the source structure from the training data. This results in better performance compared to linear inversion methods, where the prior information is limited to the 2-point covariance of the source pixels approximated with a Gaussian form, and often specified in a relatively arbitrary manner. We combine our source reconstruction network with a convolutional neural network that predicts the parameters of the mass distribution in the lensing galaxies directly from telescope images, allowing a fully automated reconstruction of the background source images and the foreground mass distribution. 
\end{abstract}

\section{Introduction}
Gravitational lensing is a powerful probe for studying many different subjects in astrophysics and cosmology, including the mass distributions in galaxies \citep[e.g.][]{Treu:04,Vegetti:12,Hezaveh:16} and galaxy clusters \citep[e.g.][and references therein]{Hoekstra:13,Natarajan:17}, the internal structures and star formation properties of galaxies at high redshift \citep[e.g.][]{Marrone:18,Spilker:18}, and the expansion rate of the universe \citep[e.g.][and references therein]{Refsdal:64,blandford:92,Suyu:14}.  An integral part of all these studies is lens modeling, through which models of both the matter distribution in the lens and the morphology of the background source are obtained. This is traditionally done using maximum likelihood (or maximum {\it a-posteriori}) methods, a procedure through which the values of the parameters describing these models are optimized by maximizing their posterior given the data.

Modeling the light distribution in the background sources requires an appropriate choice of parametrization.
A common choice is to assume that the source takes on a simple parametric form  \citep[e.g., a Gaussian or Sersic profile][]{bussmann:13,hezaveh:13b,Spilker:16}. The small number of parameters in these models are then explored using non-linear samplers such as Markov chain Monte Carlo (MCMC) samplers and are typically well constrained by the large volume of observational data.  However, for observations of complex background sources with high angular resolution and high signal-to-noise ratio (SNR), these simple parametric profiles are often found to be inadequate. Adding additional parametric source components is in general difficult since as the dimensionality of the parameter space is increased,
exploring this multi-dimensional and multi-modal space using non-linear optimizers becomes cost prohibitive \citep[although see e.g., ][]{Brewer:11}.  

In \cite{warren:03}, a method to model the background source as a pixelated image was developed, allowing a linear inversion to reconstruct the most probable image of the background source given a lens model and a particular form and strength of regularization (which becomes essential to avoid overfitting the data).  This method decomposes the modeling procedure into a nonlinear exploration of the lens parameters and a linear reconstruction of the source pixels at each step. \cite{suyu:06} formulated this method in a Bayesian framework to objectively determine the regularization strength given a fixed lens model by maximizing the Bayesian evidence.

To allow for a linear inversion, these methods assume a quadratic log-prior for the values of the source pixels. The three most commonly used forms of covariance matrices correspond to brightness, gradient, and curvature priors. These respectively enforce that either the pixel values, the gradients, or curvatures between adjacent pixels are drawn from a Gaussian distribution with mean 0 and variance given by a regularization constant.
In other words, only a two-point prior is usually imposed on sources, ignoring any knowledge of their higher order statistics. This can result in known issues, for example, allowing negative pixels or leakage of noise in the background source reconstructions and has been known to have the potential to bias the inferred parameters of the foreground structures \citep[e.g.,][]{Nightingale:18}.
The choice of pixel shape and size in the source reconstruction has also been shown to have the potential to cause systematic issues.

Moreover, the significant computational cost of performing the many matrix inversions required to find the most-probable parameters of the model in a complex, multi-dimensional parameter space makes these modeling frameworks slow and computationally expensive. Because the computational scaling of matrix inversions grows steeply with image size (roughly $t\propto N^{3}$), this expense will grow rapidly as the resolution and size of images continue to increase. For optical data the matrices are often sparse, allowing faster computations with sparse linear algebra libraries. However, for interferometric data, where the matrices are dense, these operations can be extremely costly. 

In recent years, a number of complex lens modeling tools have been developed to mitigate the above-mentioned issues \citep[e.g.,][]{Nightingale:18}. However, they typically include ad hoc procedures, for example in constructing the pixelization of the background sources or the iterative procedure through which the priors are determined. Because of their computational cost, testing for and characterizing their systematic errors is difficult and correcting them is in general non-trivial. These issues have inspired the exploration of alternative forms of parameterization for the background sources, using different basis functions, such as shapelets \citep{Birrer:15, Birrer:18}.

In this paper, motivated by the recent advances in deep learning, we investigate the use of a machine learning approach to reconstruct the images of the background sources of gravitational lenses. This is primarily driven by the fact that deep learning methods can learn complex priors from training examples \citep[e.g., ][]{Ulyanov}, circumventing the need to explicitly specify a regularization for the background sources. We explore the use of the recurrent inference machine \citep[RIM; ][]{Putzky:17}, an architecture combining convolutional neural networks  \citep[CNNs, ][]{LeCun:89} with recurrent neural networks (RNNs) to solve linear and non-linear problems iteratively.

CNNs are a class of machine learning methods that are particularly suited to image processing applications. In general, a CNN processes input images through a number of layers to produce outputs of interest.  At each layer, the image is convolved with a number of filters and acted on by a non-linear activation function. The resulting {\it feature maps} are then fed to the following layer as an input. After a number of these layers, the output of the last layer is interpreted as the output of the network. The values of the convolutional filters, also called network weights, are learned through a process known as {\it training}, where pairs of correct input-output examples are shown to the network. The values of the network weights are determined by optimizing a cost function, reducing the difference between the truth and the network's predictions. Given enough training examples, these networks can make accurate predictions on previously unseen examples using these learned parameters.

CNNs have recently seen a wide increase in use for astrophysical applications, including within the field of gravitational lensing, where CNNs have been used for performing both the tasks of lens finding \citep[e.g.][]{Lanusse:17,Jacobs:17,Petrillo:18,Pourrahmani:18,Schaefer:18}, and lens modeling \citep{Hezaveh:17,Perreault:17,Morningstar:18}, accelerating the inference of lens parameters by many orders of magnitude. 
\begin{figure}[ht]
\includegraphics[trim= 0 0 0 0,width=1\hsize]{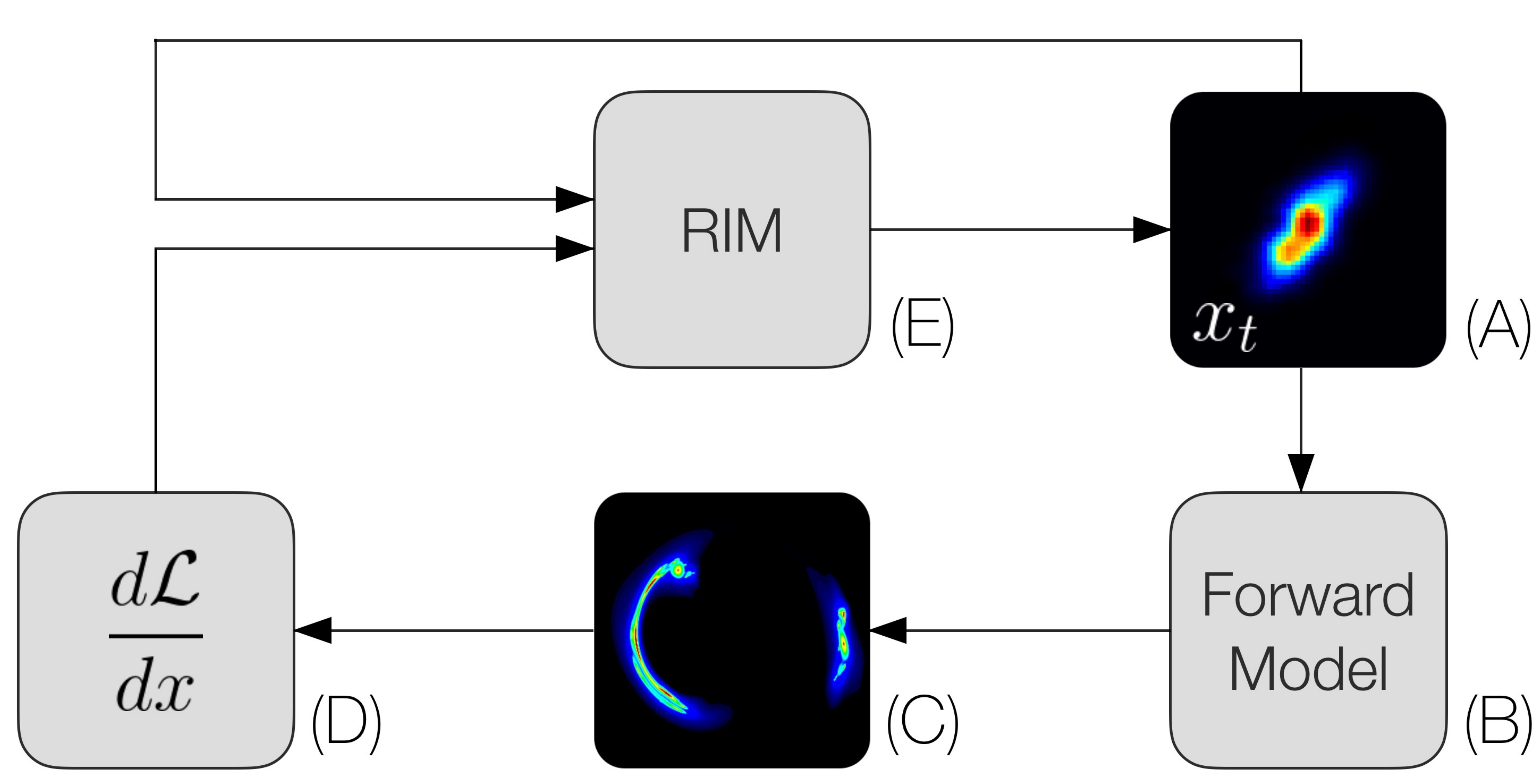}
\caption{A diagram of the structure of the reconstruction. A proposed image of the source (A) is lensed using the forward model (B, a ray-tracing simulation) to produce the lensed arcs (C, model data). The likelihood is computed by comparing the model with data. The derivative of the likelihood with respect to the values of the source pixels is calculated (D). The derivative of the likelihood and the current estimate ($x_t$) are given to the RIM (E) to produce a new estimate for the source ($x_{t+1}$).} \label{fig:RIM_diag1}
\includegraphics[trim= 0 0 0 0,width=1 \hsize]{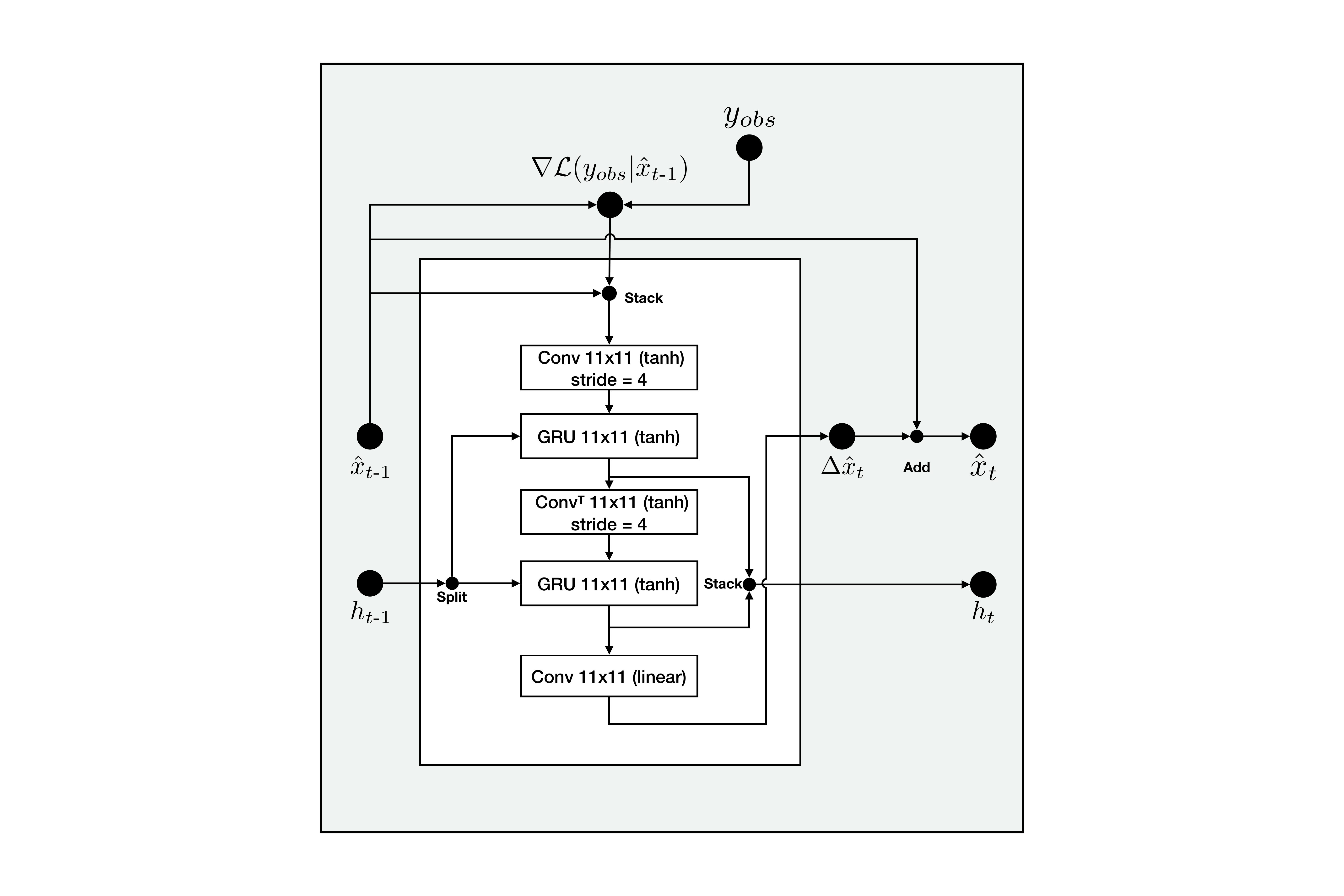}
\caption{A diagram of the RIM network cell.    The image from the previous time step ($\hat{x}_{t\text{-}1}$) is compared to the data via the likelihood (or log-likelihood).  The gradient of the likelihood is then stacked with $\hat{x}_{t\text{-}1}$ and fed to the embedding layer.  The output of the network $\Delta \hat{x}_{t}$ is added to $x_{t\text{-}1}$ to get the new prediction from the network.  The GRU cells also contain a hidden state $h$ which is updated at each time step, helping the network exhibit dynamic temporal behavior.}\label{fig:RIM_diag2}
\end{figure}

Unlike CNNs, which construct a mapping directly between a single input and output in a forward manner, RNNs operate on sequences of inputs and outputs. More specifically, they process each input through a group of layers, referred to as a {\it cell}, to both produce an output and update a {\it hidden memory state}. The next input in the series, which could sometimes be the previous output, is then processed through the same cell, which produces a new output and again updates the hidden state. This procedure can then be repeated for sequences of desired length, and through connections with the hidden state allows the network to exhibit dynamic temporal behavior. Therefore, much of the current use of RNNs is dedicated to the analysis of time series data \citep[e.g.][]{Naul:18,Charnock:17}.  However, because they are Turing-complete \citep{Siegelmann:91,Siegelmann:95} RNNs can be used to simulate algorithmic structures that perform any sequential process. 

Numerical optimization of a function can be framed as such a sequential process: starting from an initial guess, a series of steps in the parameter space of a model are taken to arrive at a final point that optimizes the value of the target function. Recently, RNN-based architectures have been used in this way to learn the process of training of other machine learning models \citep[i.e. meta-learning, ][]{LSTM-meta}. These approaches could equally be used to learn the process of optimization of other blackbox functions. The RIM \citep{Putzky:17} is an implementation of such an architecture. In \cite{Morningstar:18} we used the RIM to deconvolve dirty interferometric images prior to lensing analysis. In this work, we explore the use of this network to reconstruct the images of background galaxies from their lensed noisy data.

The outline of this paper is as follows.  We describe the network architecture along with our training set in Section~\ref{sec:methods}.  We present and discuss our results in Section~\ref{sec:results}.  In Section~\ref{sec:discussion}, we list our conclusions.

\section{methods}\label{sec:methods}

\begin{figure*}[p]
\includegraphics[width=\hsize]{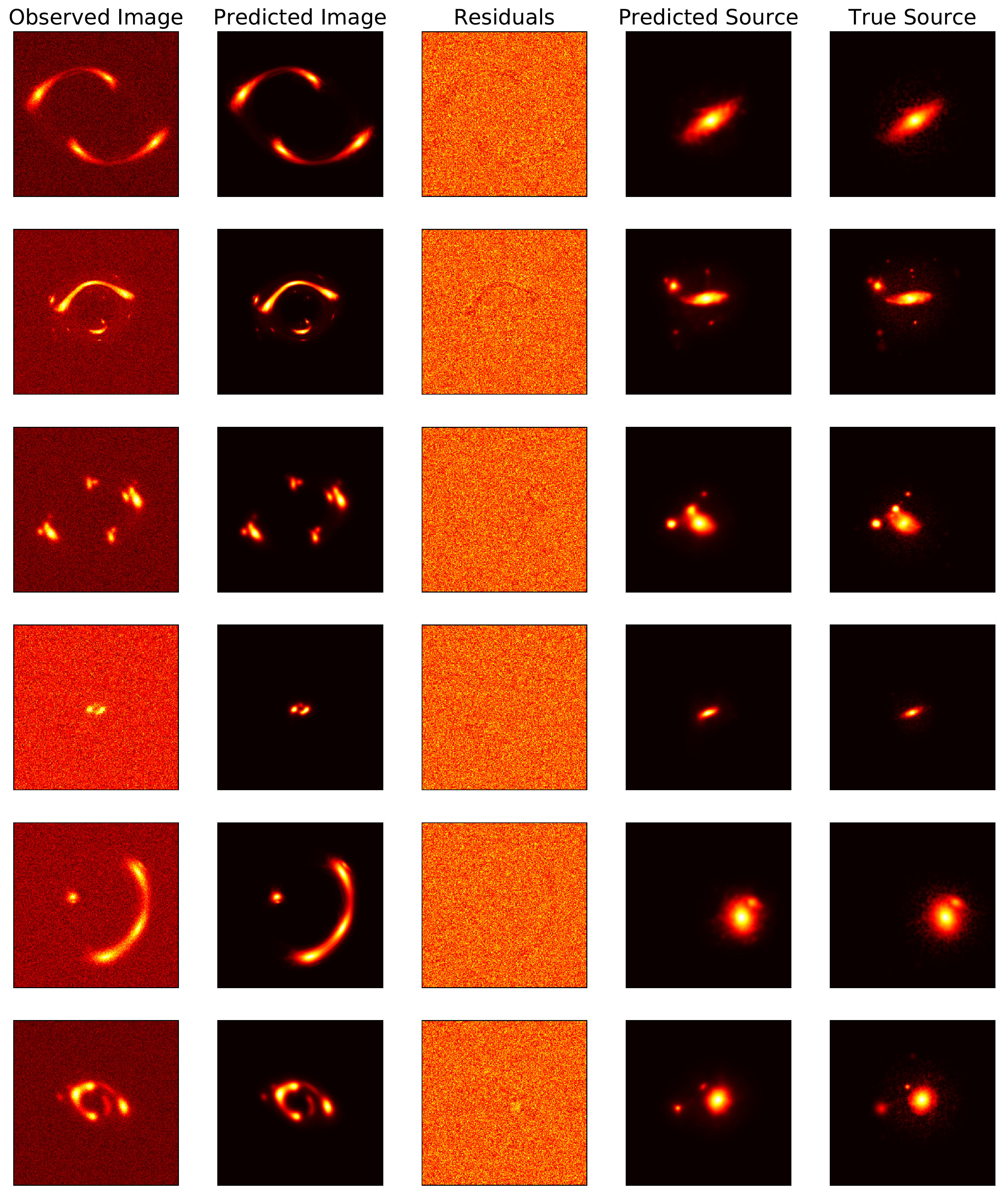}
\caption{Six example reconstructions from the test set.  The observed image (first column) is fed as an input to the RIM along with the true parameters of the lens model.  The Second column shows the forward model image created by raytracing the output of the RIM and applying smearing due to the point spread function.  The third column shows the image residuals.  The fourth column shows the reconstructed image of the background source produced at the final time step.  The fifth column shows the ground truth, for comparison.}\label{fig:RIM1}
\end{figure*}

In this section, we detail the network architecture used and describe the training set and training strategy.  
\subsection{The Recurrent Inference Machine}

The RIM framework is designed to solve linear problems of the form 
    \begin{equation}
    \label{rimeqn}
        {\bf y}=\textbf{A}{\bf x}+\mathcal{N}
    \end{equation}
for ${\bf x}$, where the vector ${\bf y}$ contains observed measurements, ${\bf x}$ is a vector of parameters that we would like to infer, $\textbf{A}$ is a known matrix mapping the parameters to the model, and $\mathcal{N}$ is a vector of noise. In the case of interest here, ${\bf x}$ represents the pixel values of the true, unlensed background source image, ${\bf y}$ is a vector of the pixel values of the observed lensed image, $\textbf{A}$ encodes the effects of lensing and observational blurring, and $\mathcal{N}$ is a vector of additive Gaussian noise. 
In this model, the likelihood of ${\bf y}$ given ${\bf x}$ is proportional to 
    \begin{equation}
    \label{RIMlikelihood}
        \mathcal{L}({\bf y} | {\bf x}) \propto \exp{\left(-\frac{({\bf y}-\textbf{A}{\bf x})^{\rm T}\textbf{C}_{\mathcal{N}}^{-1}({\bf y}-\textbf{A}{\bf x})}{2}\right)}
    \end{equation}
where the matrix $\textbf{C}_{\mathcal{N}}$ is the noise covariance matrix. Figure \ref{fig:RIM_diag1} shows a diagram of the structure of the analysis for source reconstruction.

The RIM solves equation \ref{rimeqn} recursively. At a given time step $t$, the RIM takes as input its current estimate of the true source image, $\hat{{\bf x}}_{t}$, along with the gradient of the likelihood with respect to the current estimate, evaluated at $\hat{{\bf x}}_{t}$, that is, $\left.\nabla_{{{\bf x}}}\mathcal{L}({{\bf x}})\right|_{\hat{{\bf x}}_{t}}$.  It then passes this pair through the network and outputs an update to the estimate $\Delta \hat{{\bf x}}_{t}$,  which is added to the current estimate to produce the new estimate,  
    \begin{equation}
        \hat{{\bf x}}_{t+1}=\hat{{\bf x}}_{t}+\Delta \hat{{\bf x}}_{t}\, .
    \end{equation}
The network iteratively updates its estimate of the underlying true image, using the likelihood gradient and its current prediction to guide its trajectory, in a fashion analogous to Newton's method of optimization.

Figure \ref{fig:RIM_diag2} shows a more detailed diagram of the structure of the RIM.
The network cell consists of five layers.  The first is an embedding layer that spatially downsamples the image (using a non-unity stride), but upsamples the number of features by producing more images in the channel dimension.  The second layer is the main RNN cell, which is a convolutional Gated Recurrent Unit (GRU).  The GRU shields its hidden memory state using two gates, one of which determines what to remove from the memory state (the reset gate) and the other of which determines what to add to the memory state (the update gate).  The output of the GRU is passed to a spatially upsampling layer using a transpose convolution with the same stride as the embedding layer.  The output of the spatial upsampling layer is passed to a second convolutional GRU.  The fifth layer is the output layer, which downsamples along the channel dimension.  For all layers, we use relatively large $11\times11$ filters because the images are fairly smooth on small scales.  The activation functions for the embedding, GRU, and spatial upsampling layers are all chosen to be hyperbolic tangent functions.  The GRU additionally uses sigmoid activations on its gates.  The output layer has a linear activation.  We also choose to use an output nonlinearity when generating the prediction for the source.  For this, we chose to use the sigmoid function, which has the advantage of enforcing that no output pixels have a value that is less than zero, meaning that the source can not have negative flux.  We also found that it exhibits better performance compared to a rectified linear unit nonlinearity, which would impose the same physical requirements, but which does not have a restricted upper bound.

\subsection{The Forward Model}

The RIM takes the gradient of equation \ref{RIMlikelihood}, that is, the likelihood of the trial source image ${\bf \hat{x}}$ given the observed lensed image ${\bf y}$, as an input.  To compute this likelihood, a forward, or physical, model is required.  The lens equation relates the position in the image plane to its corresponding position in the source plane
    \begin{equation}
        \beta = \theta - \alpha\, .
    \end{equation}
where $\beta$ is the position in the source plane and $\theta$ is the position in the image plane. The reduced deflection angle $\alpha$ is determined by the mass distribution in the lensing structures. 
Using the lens equation, and assuming a given interpolation scheme (typically bilinear), it is possible to construct a mapping between the source and image planes, which can be specified in matrix form.  This matrix is known as the lensing distortion matrix $\textbf{L}$ \citep[see][for how to calculate this matrix]{Treu:04}.  Using this matrix, the forward model can be constructed as follows
    \begin{equation}
        I = \textbf{B}\textbf{L}S\, ,
        \label{eq:linmodel}
    \end{equation}
where $I$ is the (noiseless) observed lensed image, and $S$ is the true background source.  The operator $\textbf{B}$ is the blurring operator, which adds the effect of smearing to the image due to the point spread function (PSF).  We can then write the log-likelihood  ($\mathcal{E}$) of the predicted source $\hat{S}$ given an observation $I_{obs}$ as 
    \begin{align}
    \label{lensingloglikelihood}
        \mathcal{E} &= -\frac{1}{2}(I_{obs}-\textbf{B}\textbf{L}\hat{S})^{T}\textbf{C}_{\mathcal{N}}^{-1}(I_{obs}-\textbf{B}\textbf{L}\hat{S}) \, .
    \end{align}
The gradient of the log-likelihood, which is used in practice by the network, is then given by
    \begin{equation}
        \nabla_{\hat{S}}\mathcal{E} = (I_{obs}-\textbf{B}\textbf{L}\hat{S})^{T}\textbf{C}_{\mathcal{N}}^{-1}\textbf{B}\textbf{L}\, .
    \end{equation}
The full input to the RIM at each time step is then obtained by stacking this with the current prediction of the source, $\hat{S}$, in the channel dimension.

\subsection{The Training Set}

The network was trained on 200 000 simulated strong lensing images. The procedure used to generate this training data is described in detail in \cite{Hezaveh:17}, but we will reiterate the major points here.  Lens models were simulated using the Singular Isothermal Ellipsoid \citep[SIE; ][]{Kormann:94} plus external shear. The SIE and shear model is composed of seven parameters that fully describe the deflections: the Einstein Radius $\theta_{E}$, the  $x-$ and $y-$components of the ellipticity $\epsilon$, the position of the center of mass of the lens ($x_{L}$,$y_{L}$), and the $x-$ and $y-$components of the shear $\gamma$. Lens models were simulated using randomly drawn values for these parameters.  

The background sources were taken to be images of galaxies from the GalaxyZoo and GREAT03 datasets.  The positions and sizes of these sources were randomly chosen, such that the minimum flux magnification found in the training set was 3.  We also imposed that the flux of the lensed sources does not fall off of the image.  The scaled images of the source were then interpolated onto a uniform grid of size $3\times3~{\rm arcsec}$ centered at the origin, so that a fixed pixel size could be used without scaling or shifting the source grid.  Both the lensed images and the background source images contain 192 pixels on each side.  The lensed images use a pixel size of $0.04~{\rm arcsec}$, and the source images use a pixel size of approximately $0.016~{\rm arcsec}$.  

Observational effects were then added randomly to the images at training time, in such a way that the network never saw the same training image twice. These effects include blurring with a PSF, approximated as a two-dimensional Gaussian function with a width randomly chosen between 0 (no PSF smearing) and $0.1~{\rm arcsec}$, and addition of noise, drawn from a Gaussian distribution with an RMS given by a uniform distribution between 0.1\% to 10\% of the peak surface brightness, resulting in an SNR of between 10 and 1000 (more heavily weighted toward low SNR).  To compute the model likelihood, both the true PSF and noise covariance matrix are provided to the forward model.

\subsection{Training}

To train the network, we use the Adam optimizer \citep{Kingma:14} with a learning rate of $2\times10^{-5}$.  This learning rate is decreased exponentially with a decay rate of 0.96 and a decay timescale of 5000 training steps.  We also employ gradient clipping in the RIM to avoid exploding gradients.  We optimize the root-mean-squared error summed over all pixels as a cost function.

In order to compute the lensing distortion matrix in the forward model, the parameters of the deflection must be provided to the network.  During training, we gave the network these parameters since they are known from the simulations.  However, when applying this technique to data this may not be possible, as the parameters of the lensing may not be known \textit{a priori}.  Therefore, we also experimented with using lens model parameters that were predicted by a feedforward convolutional neural network as described in~\citet{Perreault:17}. This network produces an estimate of the mean and standard deviation of the marginalized network posterior of the lens model parameters.  By using this to predict the lens model for the RIM, we are able to reconstruct the source without specifying any inputs other than the observed image, noise covariance matrix, and PSF.  

\section{Results and Discussion}\label{sec:results}

Once the network is trained, we test its performance on a simulated test set. For this, we use a validation set of 2000 images with background sources and lens models previously unseen by the network.  These images are each given a randomly generated PSF and noise realization drawn from the same distributions as the ones used to produce the training set.
\begin{figure}[htb]
\includegraphics[width=\hsize]{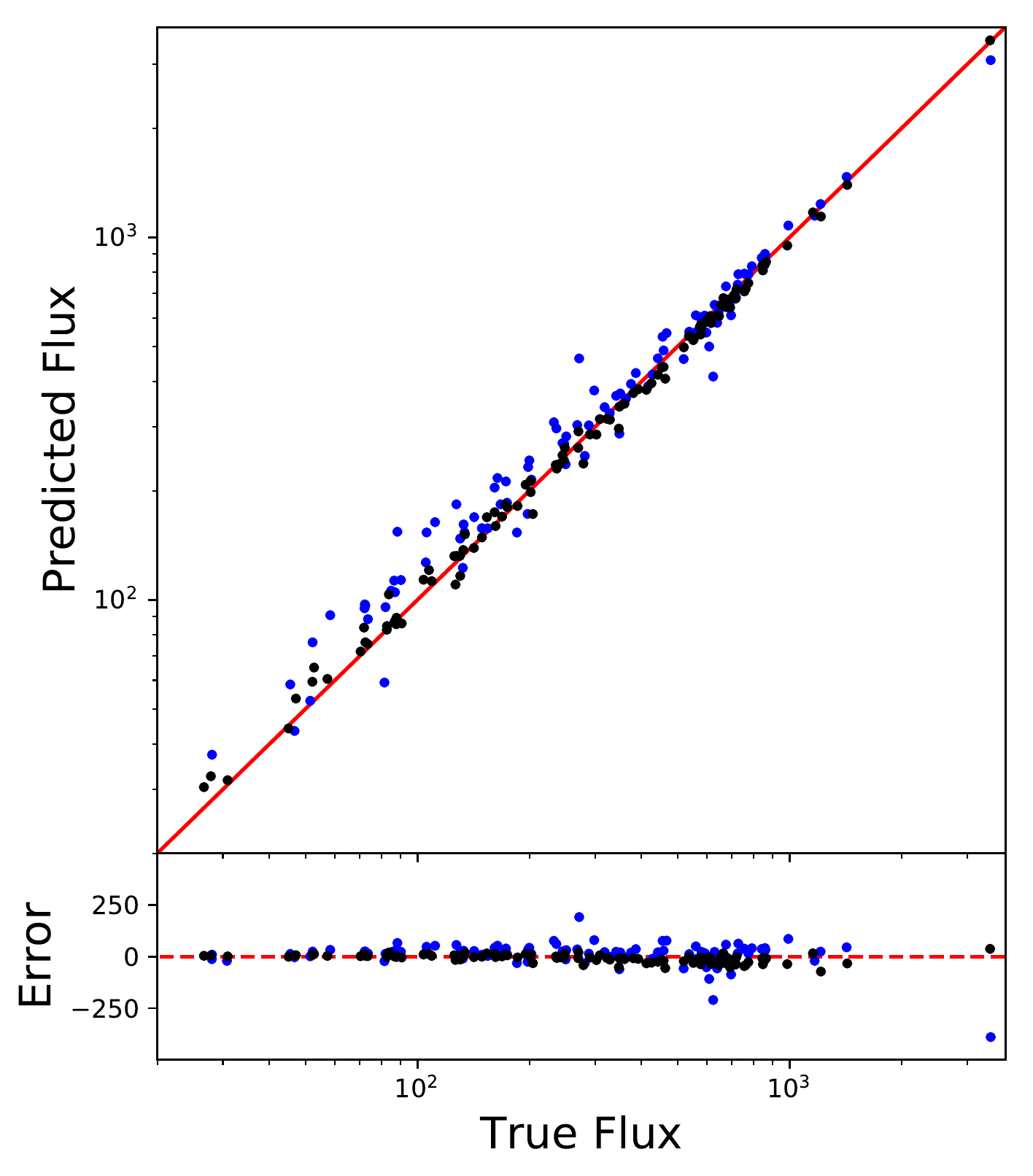}
\caption{A comparison between the predicted total flux of the background source using the RIM against the true flux of the background source.  The red line indicates a one-to-one mapping (i.e. a perfect prediction).  Black points indicate sources reconstructed using the true lens model, and blue points used a CNN to predict the lens model.  The bottom panel shows the flux error.  In both cases, the predicted flux closely corresponds to the true flux.  Units of flux here are arbitrary.}\label{fig:flux}
\end{figure}
\begin{figure}[htt]
\includegraphics[width=\hsize]{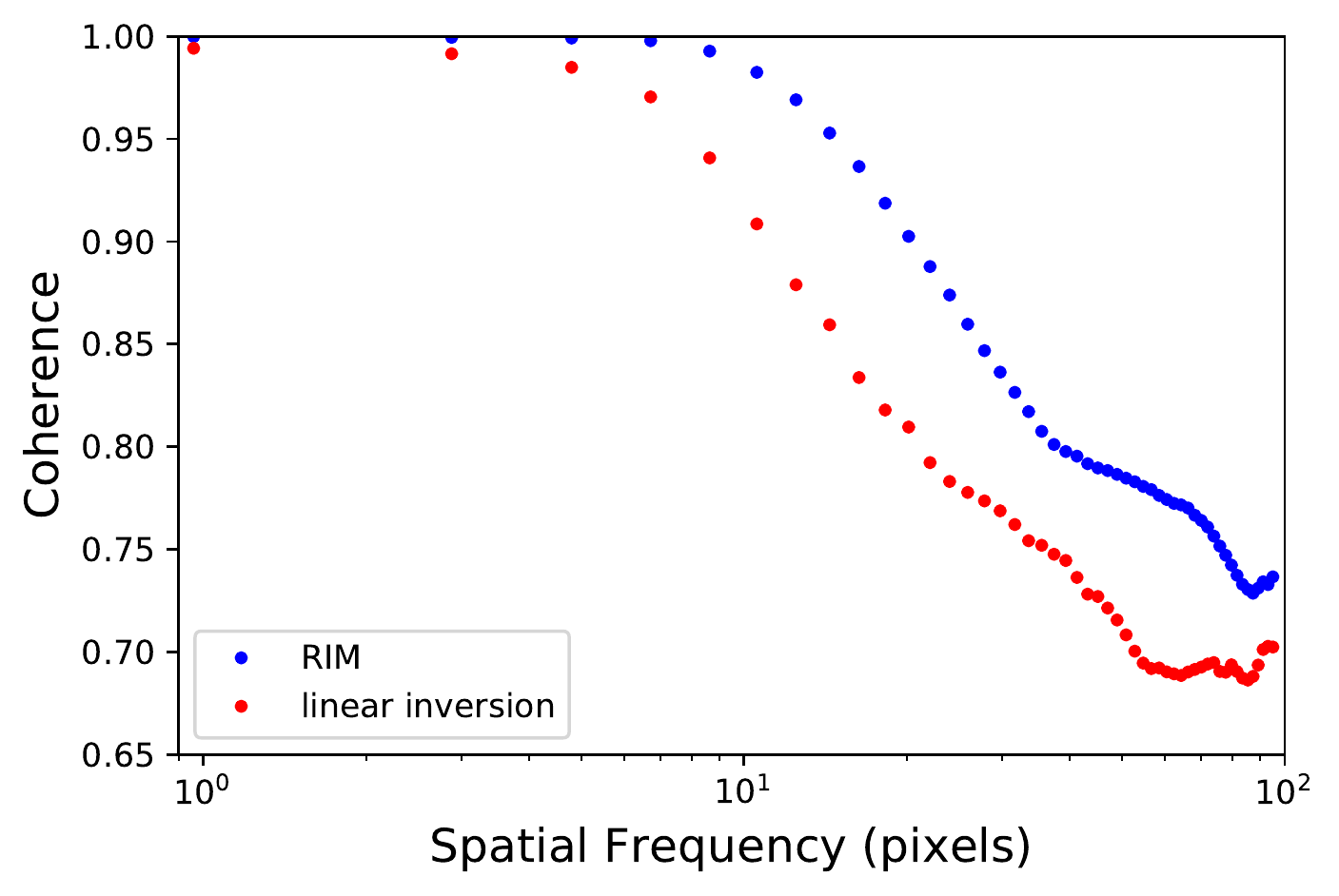}
\caption{The coherence spectrum comparing the reconstructed images to the ground truth.  The blue points show the coherence for the RIM, while the red points show the coherence for the optimally regularized semilinear inversion, computed as an average coherence at each spectral scale over all 2000 images in the test set. On large scales, the coherence is essentially unity, while on much smaller scales, the coherence drops, indicating that the model is losing information compared to the ground truth on small scales.  Over all spatial frequencies, the RIM has a higher coherence with the ground truth than the semilinear inversion.}\label{fig:coherence}
\end{figure}

\begin{figure*}[htb]
    \centering
    \includegraphics{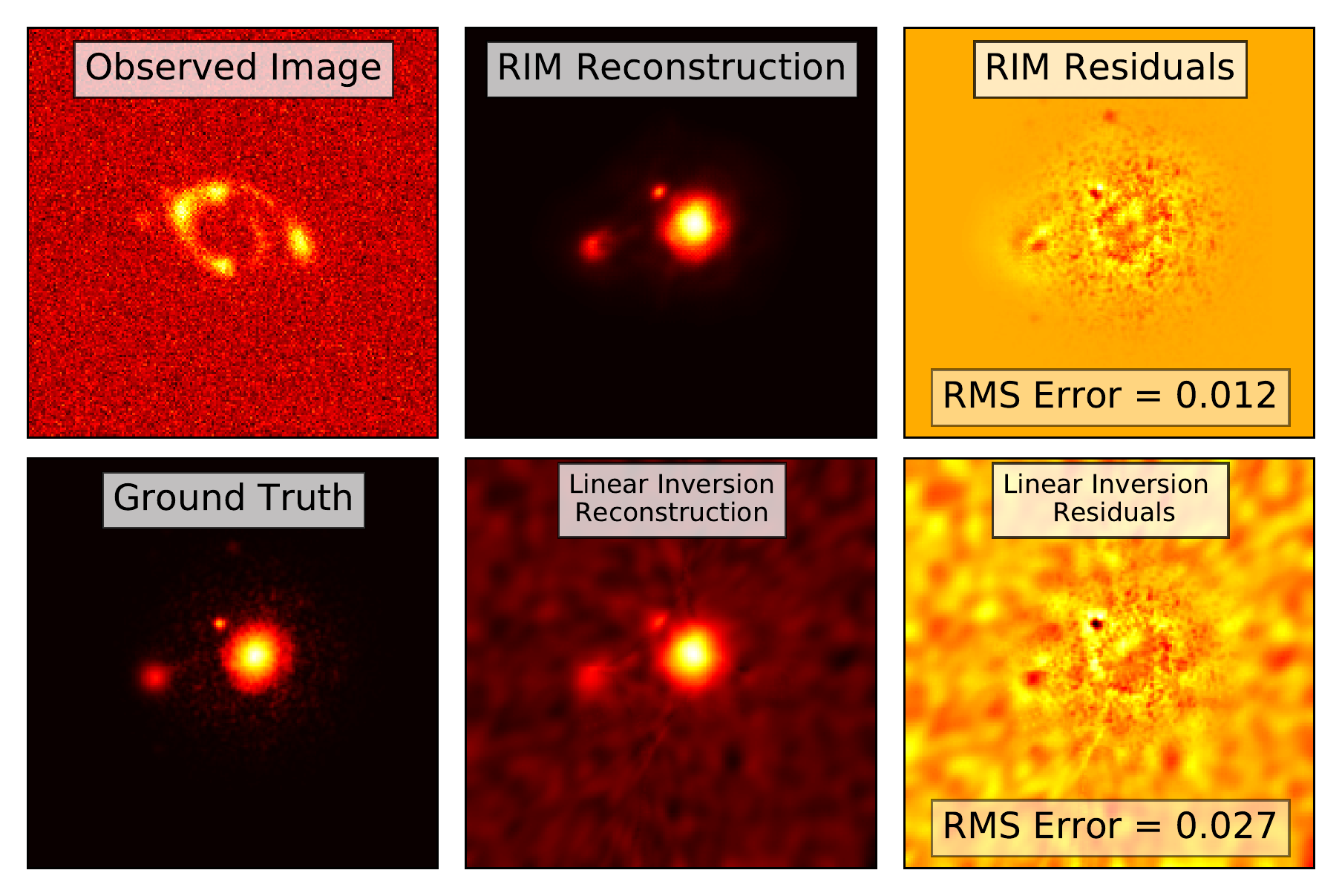}
    \caption{A comparison between the performance of the RIM and semilinear inversion on a gravitationally lensed galaxy in the validation set.  The top left panel shows the observed image, while the bottom left panel shows the true image of the background source.  The top middle and top right panels show the output from the RIM and the residuals between the RIM and ground truth.  The bottom middle and bottom right panels show the output of the semilinear inversion and the residuals.  Although the semilinear inversion was performed with the best form of regularization constant, found by maximizing the Bayesian Evidence, it still allows noise to bleed into the reconstruction, resulting in worse performance than the RIM.}
    \label{fig:RIMvsLINEAR}
\end{figure*}

We first examine the performance of the RIM in the absence of errors in the lens model.  This should provide an assessment of the peak performance of the network when the assumed distortion matrix is the correct lens mapping.
Several examples of this are shown in Figure~\ref{fig:RIM1}.  We find that the reconstructions are visually representative of the true sources and that the RMS error of the network prediction is less than 0.5\% of the peak surface brightness of the source.

As a separate metric, we compute the flux of the reconstructed sources (via a sum over all pixels) and compare this to the known flux of the ground truth images. We find that the predicted flux of the reconstructed source (determined by summing over all source plane pixels) has a median absolute error of 3\%.  The bias in recovered flux is 0.07\%, substantially lower than the random scatter. The predicted flux is shown against the true flux in Figure~\ref{fig:flux}.

\begin{figure*}[tb]
\includegraphics[width=\hsize]{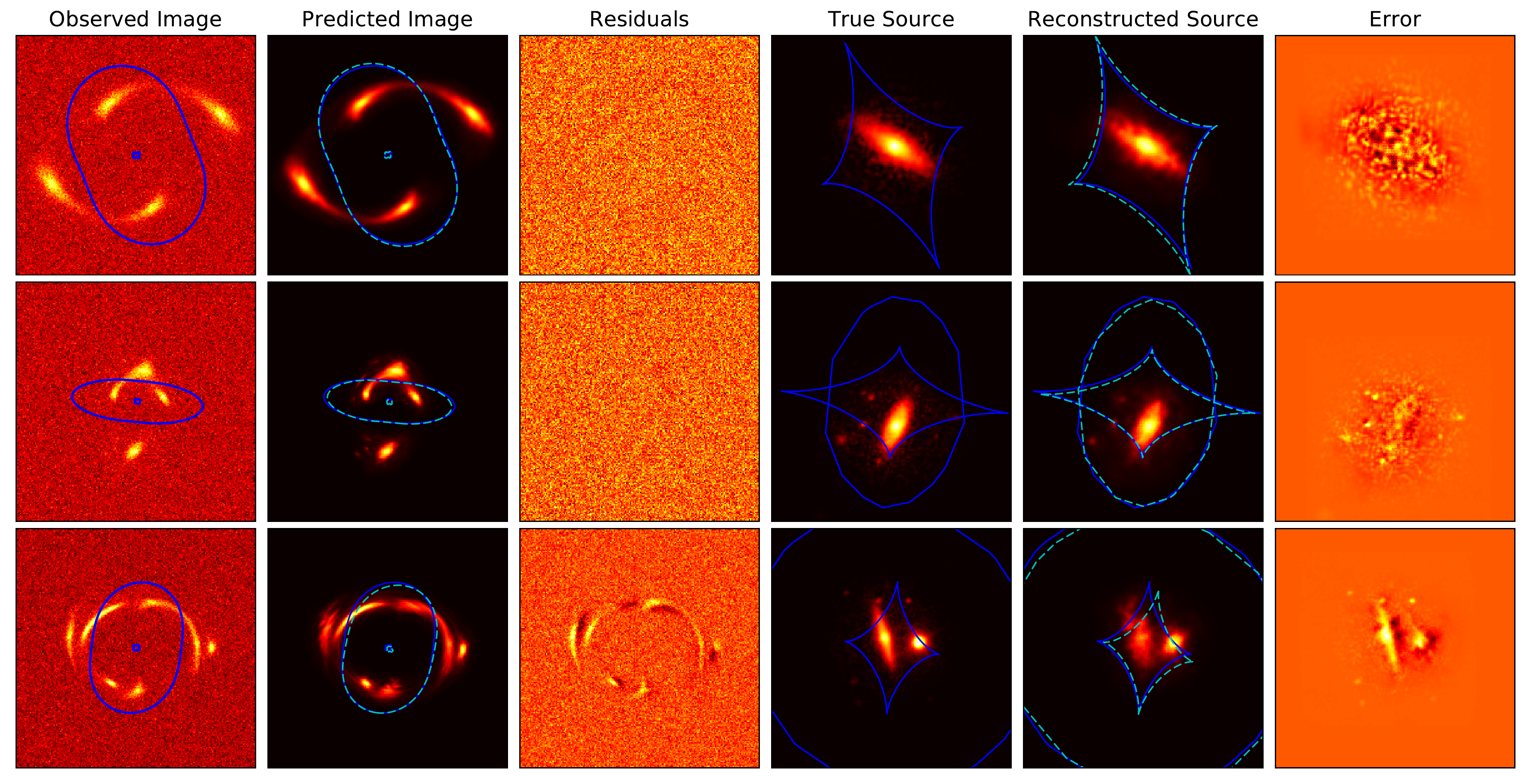}
\caption{Three example reconstructions of the background source, using a CNN to predict the lens model.  For this, we used the network from \cite{Perreault:17}.  Shown in blue are the critical curves and caustics corresponding to the true lens model.  The cyan dashed critical curves and caustics are the caustics from the predicted lens model.  In the first and second instances, the CNN accurately predicted the lens model so the caustics appear very similar.  As a result, the reconstructed source closely matches the ground truth.  In the third instance, the predicted lens model is discrepant from the truth, and so the caustics appear noticeably shifted.  The resulting errors in the mapping between the source and image plane cause a degradation in performance for this source that is easily diagnosed by inspection or by calculation of the log-likelihood.}\label{fig:RIM2}
\end{figure*}

Finally, as an additional metric of accuracy, we computed the coherence spectrum of our predicted and true images. The coherence spectrum is defined as
    \begin{equation}\label{eq:coherence}
        C(k) = \frac{P_{12}}{\sqrt{P_{11}P_{22}}} \, ,
    \end{equation}
where $k$ is the wavenumber, $P_{12}$ is the cross correlation of the images, and $P_{11}$, $P_{22}$ are the autocorrelation functions of the two images.  The coherence spectrum is used to compare the similarity between two regularly sampled signals as a function of frequency.  A coherence of $1$ implies perfect correlation between the two images, and a coherence of $0$ implies no correlation.  We plot the coherence spectrum as a function of frequency, averaged over 2000 images in our test set in Figure~\ref{fig:coherence}.  We find that our model exhibits perfect coherence for large spatial frequencies, which declines as the spatial frequency increases. For spatial frequencies of roughly $k\geq30$, our coherence is similar to that of two uncorrelated white noise realizations.  We therefore deem our model effective up to spatial frequencies of $30$.  This scale is comparable to the size of fluctuations seen in the top right panel of Figure~\ref{fig:RIMvsLINEAR}.

\begin{figure*}[tb]
    \centering
    \includegraphics[width=\hsize]{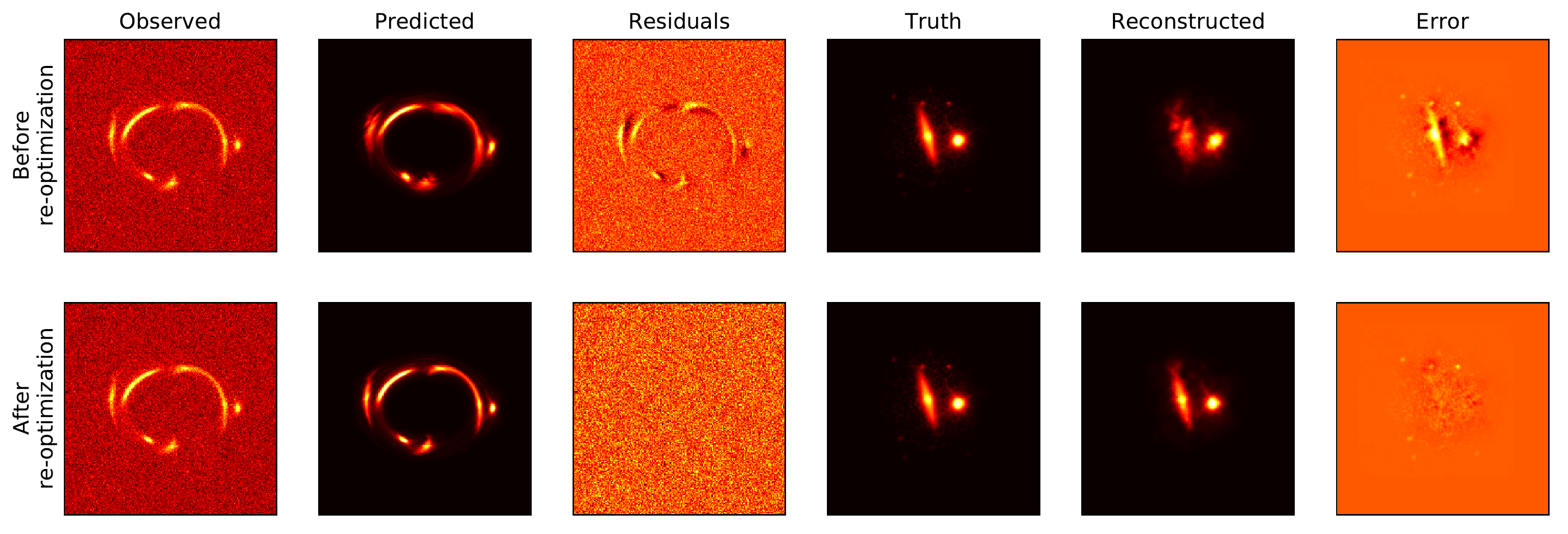}
    \caption{Results of optimization of the lens model using the RIM as a prior.  The columns are the same as in Figure~\ref{fig:RIM2}.  The example source reconstruction from the bottom row in Figure~\ref{fig:RIM2} is shown on the top row, where the RIM performs poorly due to errors in the lens model.  By optimization of the lens model using the CNN prediction as a starting position, and using the RIM to reconstruct the background source, the errors in the reconstruction can be corrected.  The corrected reconstruction is shown in the bottom row.}
    \label{fig:ErrorCorrection}
\end{figure*}

\begin{figure*}
\includegraphics[width=\hsize]{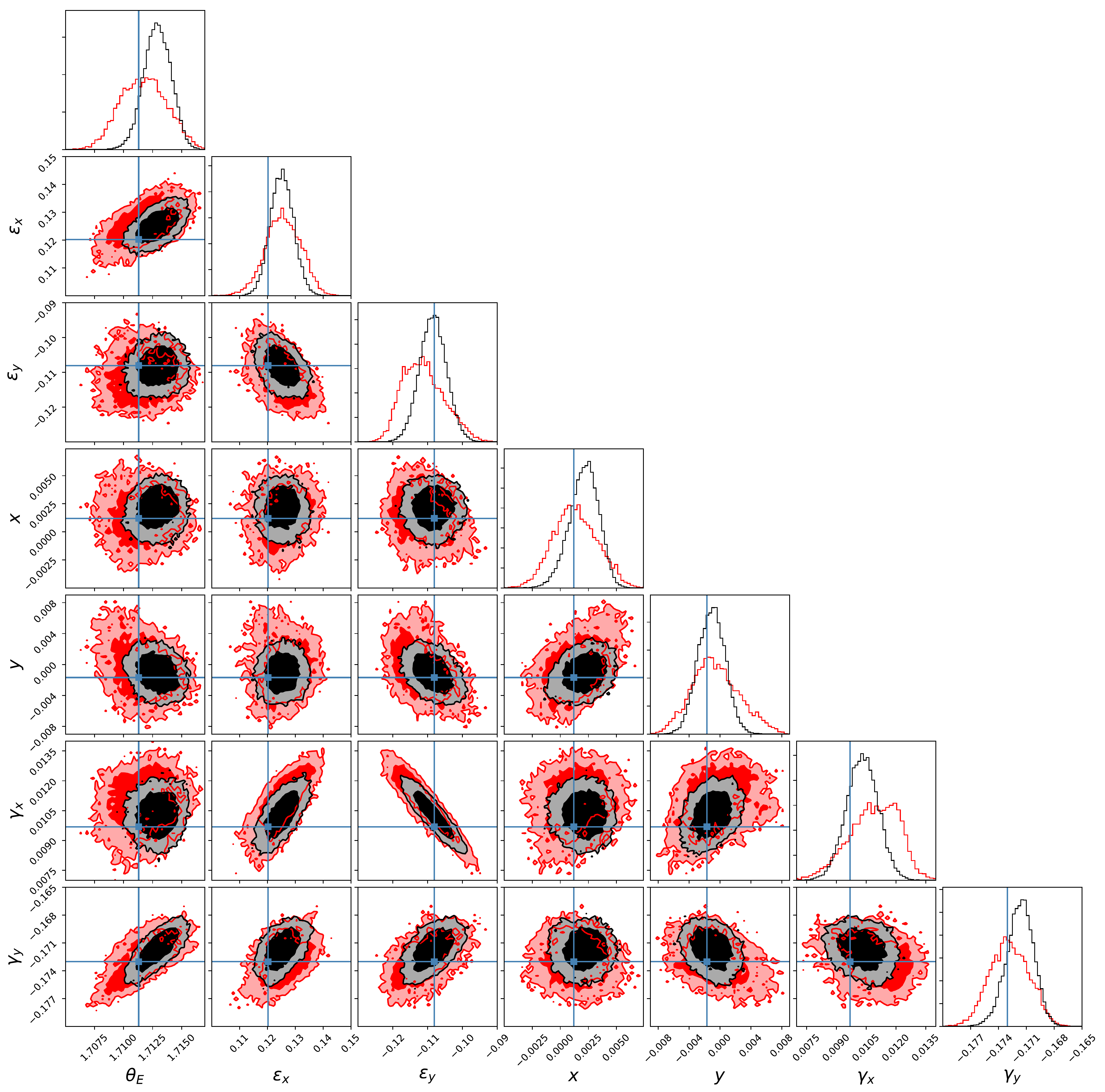}
\caption{Constraints obtained from sampling the posterior of the lens model parameters with an MCMC sampler, while performing the background source reconstructions with the RIM for one of the images in our validation set.  The true value of the lens model is shown by the blue points.  Black contours show the 68\% and 95\% confidence intervals recovered using the RIM.  The red contours show the same confidence intervals found using semilinear inversion to reconstruct the source.}\label{fig:mcmc}
\end{figure*}

We also perform a linear pixelated source inversion on a similar source grid ($192\times192$ pixels), assuming a gradient source covariance matrix. We determine the strength of the prior following the procedure described in \cite{suyu:06}. In Figure~\ref{fig:coherence} we additionally show the coherence spectrum of the reconstructed sources obtained in this way and the ground truth, averaged over all images in the test set. We note that on all scales, the RIM has a higher coherence with the ground truth than the linear inversion.  Figure~\ref{fig:RIMvsLINEAR} compares the true source morphology with both the reconstructed source obtained trough semilinear inversion and the ones given by the RIM. We note that, in linear inversions, even for the optimal strength of the regularization the resulting reconstructions contain high levels of noise, which is leaked into the source pixels with intensities as high as several percent of the peak source brightness. In addition, some pixels have negative values. This makes tasks such as estimating the intrinsic flux of the sources or analyzing their properties more difficult. The RIM reconstructions, however, succeed at suppressing the leakage of noise into the source pixels where no flux is present. We attribute this to the network learning higher order priors from the training images, which enforce zero flux over such areas.

We evaluate the log-prior of the linear model for the two reconstructions (from linear inversion and RIM) and the true background source. We find that the true source has a negative log-prior value 85 higher than the source found through linear inversion. It is not surprising that the latter has a low value, since the process of optimizing the posterior explicitly optimizes the prior. However, the fact that the true source is about 13 $\sigma$ worse than
the most-probable semilinear source suggest that the prior, even with the optimal strength, is inconsistent with the statistics of the true source. Interestingly, the source reconstructed using the RIM gives a prior value only 0.2$\sigma$ away from the true source, which suggest that it has the same statistics under this prior.

\begin{figure}[htb]
\includegraphics[width=\hsize]{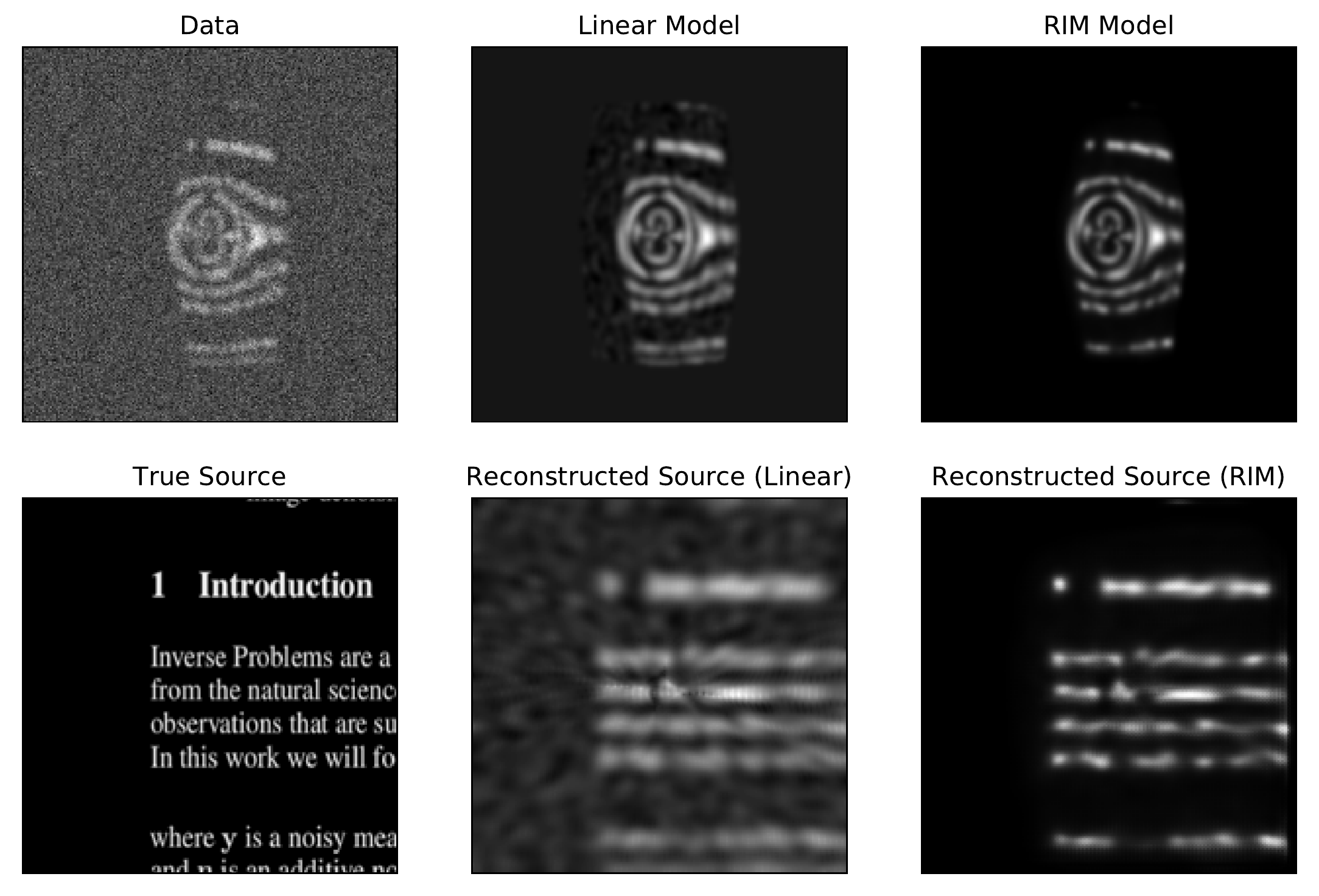}
\caption{A test of the performance of the RIM on an example far from the training data. An image of text (bottom left) is lensed and random Gaussian noise is added to it to produce simulated data (top left). Given the correct lens model, the background source is reconstructed with a linear inversion (gradient prior, bottom center) and the RIM (bottom right). The top center and top left panels show the models (lensed reconstructed source) for the linear and RIM reconstructions respectively.  Even in an example located well outside the distribution of examples provided by the training set, the RIM performs comparable to or better than linear inversion.}\label{fig:generalization}
\end{figure}

In the above discussion, it was assumed that the true parameters of the mass distribution in the lensing structures were known (these parameters were used in the forward model to predict the model observations given a proposed model of the background source, as given by equation~\ref{lensingloglikelihood}). A question of interest is if the reconstruction of the background source behaves well when a slightly incorrect lensing distortion is used. To explore this, we use the network from \cite{Perreault:17} to predict the lens model, and use these parameters to evaluate the forward model likelihood. Because the CNN introduces errors in the lens model, this can demonstrate the robustness of the RIM against imperfect foreground distortion parameters. We show three example reconstructions in Figure~\ref{fig:RIM2}.  In roughly 60\% of cases, the estimated lens model is accurate enough that the performance of the RIM is unaffected by the CNN. In the remaining cases, we find a modest decrease in performance.  The median error in flux, for example, climbs to 9\% (see Figure~\ref{fig:flux}) and the RMS pixel error climbs from 0.5\% to 6\%.  We also find that there are some (albeit rare) cases where the RIM returns morphologies that are inconsistent with the ground truth.  This occurs when the lens model parameters given by the CNN are significantly discrepant from the truth. In these cases, the failure is easily diagnosed by examining the model residuals, which are strongly inconsistent with noise, or by calculating the log-likelihood (see e.g., Figure~\ref{fig:RIM2}).

We then use a downhill optimizer to re-optimize the parameters of the foreground mass distribution (to correct the errors produced by the feedforward CNN), by minimizing the likelihood, at every step, calculating the likelihood with a background source produced by the RIM. This, in essence, becomes a traditional lens parameter optimization procedure but for the fact that the linear inversion of the source is replaced by the RIM reconstruction. We find that our optimizer converges to the true values within tens to a few hundred steps, even in cases where the initial lens model parameters were significantly discrepant from the truth (Figure~\ref{fig:ErrorCorrection}). The fact that the optimizer converges implies that the source reconstruction network is able to generalize and produce results that preserve the shape of the gradient of the likelihood in neighbourhoods around the true foreground parameters.  

This also hints at the possibility for easy automation of the lens modeling process, at least assuming a known form of parameterized model for the lensing distortion.  A simple example workflow for an automated lens modeling system is described in algorithm~\ref{alg:workflow}.  While it is a fairly unsophisticated and simple process, we found that this workflow was sufficient to achieve convergence in all of our tests.  Future works will examine generalizing this example workflow to accommodate more complex lens models, including those with multiple lens mass components.

\begin{algorithm}[H]
\caption{Example automated lens modeling workflow}
\label{alg:workflow}
\begin{algorithmic}[1]
\Procedure{Automated lens modeling}{}\newline
\textbf{Input:} Image of gravitational lens, model of instrumental PSF, and expected noise rms in each pixel \newline
\textbf{Output:} Reconstructed image of the background source, and a parameterized model for the lensing distortion
\State{Predict lens model using feedforward CNN} 
\State{Reconstruct background source with predicted lens model using RIM}
\While{the residuals are not consistent with noise}
    \State{Optimize lens model with non-linear optimizer}
    \State{Reconstruct the source with the optimized lens model using RIM}
\EndWhile
\EndProcedure
\end{algorithmic}
\end{algorithm}

We then explore if it is possible to sample the posterior of the lens model parameters with an MCMC sampler, while performing the background source reconstructions with the RIM. We use the affine-invariant algorithm of \cite{mcmchammer}, as implemented in \cite{emcee} to sample the posterior of the lens parameters. Figure \ref{fig:mcmc} shows the corner plot of the samples. We find that the sampled parameters are well constrained and relatively Gaussian, and that the procedure can identify known degeneracies in the lens model (e.g., the degeneracy between ellipticity and external shear), again suggesting that the network can generalize to examples beyond (but close to) its training data, where the lens models are slightly incorrect. We also show the constraints on the lens model obtained using semilinear inversion to reconstruct the source.  While both achieve results consistent with the ground truth, the RIM achieves higher precision. We attribute this to the more constraining prior learned by the network.

In this work, we only obtained a point estimate for the pixelated morphologies of background sources without obtaining their uncertainties. The uncertainties of the predictions of neural network is an active area of research. Previously in \citet{Perreault:17} we demonstrated the use of variational inference in obtaining approximate uncertainties for these predictions. In the future we plan to extend this method to the predictions of the RIM to obtain an estimate of the uncertainties of the predictions.

Machine learning systems are often unable to deal with generalization to different tasks.  In particular, feed-forward CNNs perform substantially worse when the test data distribution is different from the training data distribution.  Because we use images of particular subsets of galaxies from the GalaxyZoo and GREAT3 challenge data, it is possible that the galaxies that make up the background sources of real gravitational lenses possess different morphologies than the galaxies used in our training set.  However, we speculate that the differences between the training and test distributions matter less for the RIM than they do for a typical feed-forward CNN, because it learns a procedure to optimize the likelihood given a set of observed data.  Therefore, the RIM may still perform adequately, even if its test data are significantly different than the training data.  To test this hypothesis, we examined the performance of the network on lensed images, where the background source was an example far from the training data. For this, we chose to use an image of text. Figure \ref{fig:generalization} shows the results of this experiment. The top left panel shows the data (lensed noisy image of a text). The right and middle lower panels show the reconstructions using linear inversion and RIM respectively. Remarkably, we found that RIM performed reasonably well even though the local and global structures of the text image are very different that the unlensed galaxies used for training. The reconstruction identifies the positions of letters, but due to smearing by the PSF, it appears to be unable to produce a legible image.  We have confirmed this result using images of handwritten numbers from the MNIST dataset \citep{LeCun:98}. Even in a regime outside of its training set, the RIM appears to perform well compared to a linear inversion.

Since the only input of the network is the gradient of the log-likelihood with respect to the source pixels, one can easily generalize the application of this method to interferometric data by simply modifying the forward model to predict the model visibilities and use them to calculate the likelihood in the $uv$-space. This is essentially equivalent to replacing matrix $\textbf{B}$ in equation \ref{eq:linmodel} with a matrix performing the Fourier transform. However, since only a forward prediction is required (i.e. the prediction of visibilities given a lensed source images is needed), this could be done using a fast Fourier transform (FFT). Given the extreme computation of cost of linear source inversions for large visibility sets, this could result in many orders of magnitude improvement in speed and computational cost.

\section{Conclusions}\label{sec:discussion}

We present a method that uses recurrent convolutional neural networks to recover the morphology of the background sources of gravitational lenses from telescope data.  From our tests of this method we draw the following conclusions:
\begin{enumerate}
    \item We found that the reconstructed sources predicted by neural network exhibit better fidelity to the ground truth images (measured using both mean-squared error and the coherence spectrum) than linear inversion methods.
    
    \item We showed that neural networks are able to reconstruct the image of the background source using only the image, the noise covariance matrix, and the point spread function; information easily discerned from the data itself.  To perform these source reconstructions, we use a convolutional neural network to predict the parameterized form of the mass distribution in the lensing structures from the observed images.
    
    \item We observed that small errors in the lens model can occasionally cause errors in the source reconstruction when predicting the lens model using a CNN.  The errors on both the lens model and the source morphology can be corrected by modeling the image in a maximum likelihood fashion, using the RIM as a constraint on the source morphology.
    
    \item We used our network as a source reconstruction module to sample the parameters of the mass distribution in the lensing structure using an MCMC procedure.  The estimated parameters and uncertainties exhibit fidelity to the ground truth. We also sampled this posterior using a semi-linear modeling method. The joint probability density of the parameters obtained from these two procedures exhibit similar degeneracies, however the RIM resulted in higher precision compared to semilinear models.

    \item We tested the performance of our network on examples outside the training data, by both providing the network incorrect lens models and also requiring the networks to perform reconstructions of images of text. We found that the networks performed well in both instances, showing that they can potentially be robust against possible discrepancies between training and test data sets.
    
    \item Our current work does not provide an estimate of the uncertainty of the reconstructions. In future works, we plan to investigate methods for obtaining the uncertainties of these reconstructions in a manner similar to the work presented in \citet{Perreault:17}.

\end{enumerate}

\bibliographystyle{yahapj}

\end{document}